# Trapping light by mimicking gravitational lensing

C. Sheng[1], H. Liu[1,*], Y. Wang[1], S. N. Zhu[1], and D. A. Genov[2]

[1]National Laboratory of Solid State Microstructures & Department of Physics, National Center of Microstructures and Quantum Manipulation, Nanjing University, Nanjing 210093, People's Republic of China

[2]College of Engineering and Science, Louisiana Tech University, Ruston, Louisiana 71270, USA

*Corresponding author: liuhui@nju.edu.cn

**One of the most fascinating predictions of the theory of general relativity is the effect of gravitational lensing, the bending of light in close proximity to massive stellar objects. Recently, artificial optical materials have been proposed to study the various aspects of curved spacetimes, including light trapping and Hawking's radiation. However, the development of experimental 'toy' models that simulate gravitational lensing in curved spacetimes remains a challenge, especially for visible light. Here, by utilizing a microstructured optical waveguide around a microsphere, we propose to mimic curved spacetimes caused by gravity, with high precision. We experimentally demonstrate both far-field gravitational lensing effects and the critical phenomenon in close proximity to the photon sphere of astrophysical objects under hydrostatic equilibrium. The proposed microstructured waveguide can be used as an omnidirectional absorber, with potential light harvesting and microcavity applications.**

During the total solar eclipse in 1919, Arthur Eddington and collaborators performed the first direct observation of light deflection from the Sun, which validated the Einstein's theory of general relativity[1]. According to Einstein's theory, the presence of matter energy results in curved spacetime and the complex motion of matter and light along geodesic trajectories that are no longer straight lines. The general theory of relativity has been highly successful, with many of its predictions validated, including the precession of Mercury[2], gravitational time dilation and gravitational red shift[3], expansion of the universe[4], and frame-dragging[5]. On the other hand, given the analogy between the macroscopic Maxwell's equations in complex inhomogeneous media and the free-space Maxwell's equations on the background of an arbitrary spacetime metric[6-12], the study of light propagation in artificially engineered optical materials and the motion of massive bodies or light in gravitational fields are closely related. The invariance of the Maxwell's equations under coordinate transformations has been used to design invisibility cloaks at microwave and optical frequencies[13-19] as well as a vast variety of transformation optical devices[20-23]. Furthermore, this analogy has been proposed to mimic well-known phenomena that are related to general relativity but are difficult to directly observe using existing astronomical tools. In particular, the transformation optics approach can be used to mimic black holes[24-28], Minkowski spacetimes[29], electromagnetic wormholes[30], cosmic strings[31], the Big Bang and cosmological inflation[32,33], as well as Hawking radiation[34,35]. Although the theoretical foundations behind the design of "toy" models of general relativity are now sufficiently well understood, the experimental validation of the transformation optics approach to mimic the scattering of electromagnetic radiation in close proximity to actual celestial objects remains a challenge, especially for visible light.

In this work, we propose a flexible experimental methodology that allows the direct optical investigation of light trapping around a microsphere, which simulates the gravitational lensing due to power law mass–density/pressure distributions, including light trapping at the photon sphere similar to that around compact neutron stars and black holes (Fig. 1a). Our approach uses a microsphere that is embedded into a planar polymer waveguide (Fig. 1b) that is formed during a controlled spin-coating process. Given the surface tension effects, the waveguide around the microsphere is distorted resulting in a continuous change in the waveguide effective refractive index, which under certain conditions can mimic the curved spacetimes caused by strong gravitational fields. Based on direct fluorescence imaging, we observe the lensing and asymptotic capture of the incident light at an unstable circular orbit that corresponds to the photon sphere of a compact stellar object. These observations clearly demonstrates that the proposed experimental methodology provides a useful "toy" model to study both near- and far-field electromagnetic effects under a controlled laboratory environment, similar to the gravitational lensing described in general relativity. The experimental observations are in excellent agreement with the developed theory, validating both the derived exact solution of the Einstein field equations and the predicted geodesic trajectories of massless particles in the inherent curved spacetime.

**Results**

**Sample fabrication and optical characterization.** A structured waveguide is fabricated in the experiment (Fig. 1b). During the fabrication process, a 50 nm thick silver film is initially deposited on a silica ($SiO_2$) substrate. A grating with a period of 310 nm is drilled across the silver film using a focused ion beam (FEI Strata FIB 201, 30 keV, 11 pA). A polymethylmethacrylate (PMMA) resist is mixed with oil-soluble CdSe/ZnS quantum dots (at a volume ratio of 2:1). A powder consisting of microspheres 32 μm in diameter is then added to the mixture. The CdSe/ZnS quantum dots are added for the purpose of fluorescence imaging, while the microspheres provide the important functionality of creating a gradual change in the PMMA thickness (in close proximity to the microspheres) which results in a gradient index waveguide. The solution is deposited on the silver film through a spin-coating process, and the sample is dried in an oven at 70 °C for 2 h. In the process, the thickness of the PMMA layer can be controlled by varying the spin rate, evaporation rate, and solubility of the PMMA solution. In the zone located far from the microsphere, the PMMA layer is uniformly thick (approximately 1.0 μm). In the region near the microsphere, the waveguide thickness gradually increases due to surface tension effects before and during the baking process. This phenomenon is indirectly observed through the interference pattern (Fresnel zones) around the microsphere (Fig. 2a), where the sample is illuminated with a white (top) and blue (bottom) light. The interference minima/maxima depend on the PMMA thickness and can be used to extract the thickness profile as a function of the distance to the center of the microsphere. In addition, the surface profile of the PMMA layer is directly measured using atomic force microscopy (AFM). The measured thickness profile (Fig. 2b) is in good agreement with the results retrieved from the interference measurements.

The light deflection provided by the variable thickness waveguide and observed in proximity to the microsphere can be described using the waveguide effective refractive index. In the experiment, the structured waveguide consists of an air/PMMA/silver/SiO$_2$ multilayer stack (Fig. 1b) and can be considered as step-index planar waveguides. The dispersion relationship of the waveguide transverse magnetic (TM) modes (see Supplementary Information) is used to extract the effective refractive index around the microsphere which is depicted in Fig. 2c. The waveguide index (for the $TM_{0,6}$ mode) rapidly decreases with the distance from the microsphere according to a power law dependence $n_e^2 \approx n_{e,\infty}^2 \left[ 1 + (a/r)^4 \right]$, where the parameters $a = 28.5 \mu m$ and $n_{e,\infty}^2 = 1.1$ correspond to the best fit. Within the immediate proximity to the microsphere, the refractive index approaches the bulk values of the PMMA, $n_{PMMA}^2 = 2.31$.

To study the ray propagation in close proximity to the microsphere, a 405 nm light from a CW laser is coupled into the waveguide through a grating (Fig. 1b). As the coupled light propagates within the waveguide, it excites the quantum dots that then reemit at 605 nm. The fluorescence emission from the quantum dots is collected by a microscope objective (Zeiss Epiplan 50×/0.17 HD Microscope Objective) and delivered to a charge-coupled device camera. The obtained fluorescence image is then used to analyze the ray trajectory. One particular example is given in Fig. 2d. The incident light is deflected as it passes in the vicinity of the microsphere. In the following section, this phenomenon is revealed to be closely related to the deflection of light in a centrally symmetric curved spacetime that correspond to degenerated fluid in hydrostatic equilibrium with asymptotically polytropic equation of state.

The complete range of optical phenomena associated with the proposed microstructured waveguide is mapped by a set of measurements in which the excitation point is gradually moved along the grating and toward the microsphere. In the supplementary materials, a movie file is provided showing the tuning process used in the experiment[36]. The change in the fluorescence pattern, which captures the interaction of the incident beam with the inhomogeneous effective refractive index of the waveguide, is presented in Fig. 3a. We observe a gradual increase in the light deflection as the distance to the microsphere decreases. Due to the finite excitation spot size, which corresponds to a Gaussian beam waist size ($\sigma \approx 3\mu m$), the beam fans out with the outside beam envelope deflected at lower angles. For excitation with an impact parameter, i.e., the perpendicular distance between the beam and the center of the microsphere, that approaches a critical value $b_c \approx 39 \mu m$, the impinging light approaches an unstable photon orbit, i.e. photon sphere, at a radius of $r \approx a$. The photon sphere splits the entire space into two domains such that if the impact parameter is larger than the critical the impinging light approaches the microsphere until it reaches a point of closest approach (turning point) and is then deflected back into space while for impact parameters less than the critical the light is captured. To validate our experimental finding, full-wave finite difference calculations are performed using finite difference COMSOL Multiphysics software. The theoretically obtained scattering profiles (Fig. 3b) are nearly identical to the experimental data. The two main experimental finding are again observed: (i) the deflection angle increases with decreasing impact parameter, and (ii) the impinging light is captured by the system for impact parameters below the critical value $b \leq b_c$. These results are reminiscent to the gravitational lensing as well as the existence of a photon sphere around stellar objects such as ultracompact neutron stars and black holes[37]. Hence, our system may provide a useful "toy" model to study the electromagnetic scattering and light capture due to such unique astrophysical objects.

**Discussions**

**Curved spacetimes around the microsphere.** To validate the proposition from above, we consider static centrally symmetric spacetimes described by the isotropic metric; $ds^2 = -g_{00}(r)dt^2 + g_{rr}(r)d\vec{x}^2$. The metric must be a solution of the Einstein field equations $G_{uv} = -T_{uv}$, with a stress-energy tensor $T_{uv} = \rho u_u u_v + p(u_u u_v - g_{uv})$ that depends on centrally symmetric mass-density $\rho$ and pressure $p$ distributions. The field equations can be solved either by providing the equation of state $p = p(\rho)$, or by using a generating function[37, 38]. Here, we rely on the later approach by enforcing the matching condition, i.e. the effective refractive index of the metric $n = \sqrt{g_{rr}/g_{00}}$ [25], must coincide with that of the experiment and then proceed to obtain the unknown metric elements, mass-density and pressure. The field equations are found to have an exact solution with a metric

$$ds^2 = \frac{A}{r^4 n^4(r)} \csc^2\left(\sqrt{\frac{5}{2}} \cot^{-1}(r^2)\right)\left[-dt^2 + n^2(r)d\vec{x}^2\right] \quad (1)$$

where $A$ is an integration constant (see Supplementary Information). The metric Eq. 1 is finite and corresponds to asymptotically flat free space at large distances. The required pressure and mass-density are also obtained showing an asymptotic (for $r > a$) equation of state $p = k\rho^{1+1/n}$ with a polytropic index $n = \frac{3}{2}$. A large variety of gravitational objects in hydrostatic equilibrium can satisfy such an equation of state, including degenerate star cores such as those in neutron stars, red giants and white dwarf, and non-isothermal gas clouds with an interior that is cooler than the exterior [37].

The motion of a light ray in the curved spacetime Eq. 1 is described by the Largangian $\mathcal{L} = (1/2)(ds/d\tau)^2$ where $\tau$ is the trajectory parameter. The Euler-Lagrange equations are then solved giving an explicit solution for the ray trajectories as function of the azimuthal angle $\varphi$ in the form

$$u(\varphi) = u_0 + u_t sn\left(\frac{q(\varphi - \varphi_0)}{u_t}\middle| u_t^4\right) \quad (2)$$

where $u = a/r$ is the inverse radial coordinate, $\varphi_0$ is the angle of incidence, $u_0$ is the initial position, and $sn$ is the Jacobi elliptic function. The solution depends on the external turning point $u_t = a/r_t = (b/b_c)\left(1 - \sqrt{1 - (b_c/b)^4}\right)^{\frac{1}{2}}$, i.e. the position of closest approach. Clearly, for in-falling rays a turning point exists only if the impact parameter is larger than the critical value $b \geq b_c = a\sqrt{2}$, otherwise the rays will be captured within the spatial domain below the photon sphere $r \leq a$. Thus, our system can be described with a total capture cross-length $\sigma_c = 2b_c = 2a\sqrt{2}$, indicating that any light ray that approaches the microsphere within such a spatial range will be captured. The capture cross-length is independent on the direction from which the light has been emitted which exemplifies the omnidirectional properties of the system and points toward possible application in light steering and energy harvesting devices. Finally, the total deflection angle for in-falling rays with $b \geq b_c$ is obtained from Eq. 2 as:

$$\theta = 2K\left[u_t^4\right]\sqrt{1 + u_t^4} - \pi \quad (3)$$

where $K$ is the complete elliptical integral of the first kind. If the incident ray traverses the spatial domain away from the photon sphere ($b \gg b_c$), then $\theta \to 3\pi(b_c/2b)^4$, and an inverse power law dependence of the deflection angle with the impact parameter is observed. This indicates that objects described with the metric Eq.1 will exhibit gravitational lensing with an equivalent lens equation of the form, $1/s_1 + 1/s_2 = 1/f$, where the "focal" length $f = 4a/3\pi u_t^5$ depends on the distance of the closest approach $r_t = a/u_t$ commonly referred to as the Einstein ring radius; and $s_1$ and $s_2$ are the distances to the source and image, respectively.

The comparison between the experimentally measured and theoretically obtained deflection angles are shown in Fig. 4. Given that the incident beam in the experiment has a finite size, two deflection angles can be unambiguously extracted from the experiment, particularly those that correspond to the beam envelope impact parameters $b_\pm = b_0 \pm \sigma$, where

$b_0$ is the impact parameter that corresponds to the maximum beam intensity. The deflection angles are then plotted versus the geometric average of the two distances of the closest approach (inset in Fig. 4). The experimental data are consistent with the theoretical findings, indicating that our experimental method can describe both the far-field scattering and the critical behavior close to the photon sphere related to the curved spacetimes given by Eq. 1. Aside from light deflection, the gravitational time delay (or Shapiro effect) may also be investigated using our experimental setup. We must note that within extended stellar objects apart from the gravitational effects the light rays will also be affected by the object material constituents (change particles, atoms and molecules). These types of scattering processes while important are rather complex in nature and go beyond the scope this work which only aims at investigating the effects of gravity.

In conclusion, we have experimentally demonstrated an optical analogue of the effects of gravity on the motion of light rays, including light deflection, Einstein rings and photon capture. The "gravitational field" effect is achieved using an inhomogeneous effective refractive index provided by a microstructured waveguide spin-coated in the presence of a microsphere. The deflection and capture of light are directly observed based on the fluorescence imaging method. An exact solution of the Einstein field equations is obtained showing that the proposed "toy" model can mimic the effect of gravity due to spherically symmetric object in hydrostatic equilibrium with asymptotically polytropic equation of state. Our method may also be applied to control light propagation in integrated optoelectronic elements, light splitters and benders, omnidirectional absorbers and energy harvesting devices.

**Acknowledgements**

This work has been supported by the National Key Projects for Basic Researches of China (No. 2012CB933501, 2010CB630703 and 2012CB921500), the National Natural Science Foundation of China (No. 11074119, 60990320 and 11021403), the Louisiana Board of Regents and NSF under Contracts No.LEQSF (2007-12)-ENH-PKSFI-PRS-01, No. LEQSF (2011-14)-RD-A-18, the Project Funded by the Priority Academic Program development of Jiangsu Higher Education Institutions (PAPD), New Century Excellent Talents in University (NCET-10-0480), the doctoral program(20120091140005) and Dengfeng Project B of Nanjing University.

**Author contributions**

C. S., H. L., Y. W.  and S. N. Z. proposed and carried out the experiment; D. A. G. contributed to the experimental characterization and interpretation, proposed and developed the theory; D. A. G. , C. S., and H. L. co-wrote the manuscript.

**Additional information**

The authors declare no competing financial interests. Supplementary information accompanies this paper at www.nature.com/naturephotonics. Reprints and permission information is available online at http://www.nature.com/reprints. Correspondence and requests for materials should be addressed to H.L.

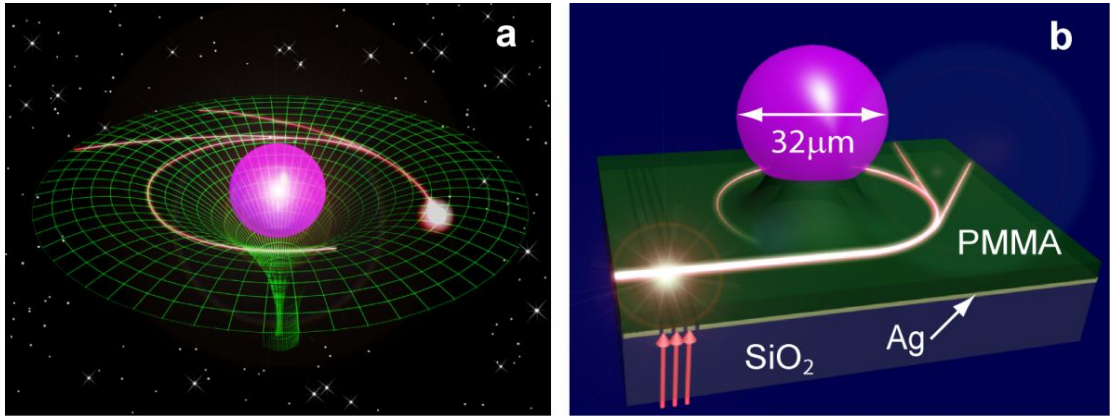

**Figure 1. Analogue of light deflection in gravitation field and microstructured optical waveguide.** (a) Depiction of light deflection by the gravitational field of a massive stellar object. (b) The schematic view of the microstructured optical waveguide formed around a microsphere and used to emulate the deflection of light by a gravitational field. In the experimental setup, a grating is drilled across a 50nm thick silver layer which is then used to couple the incident laser light into the waveguide. The red arrows denote the incident laser beam.

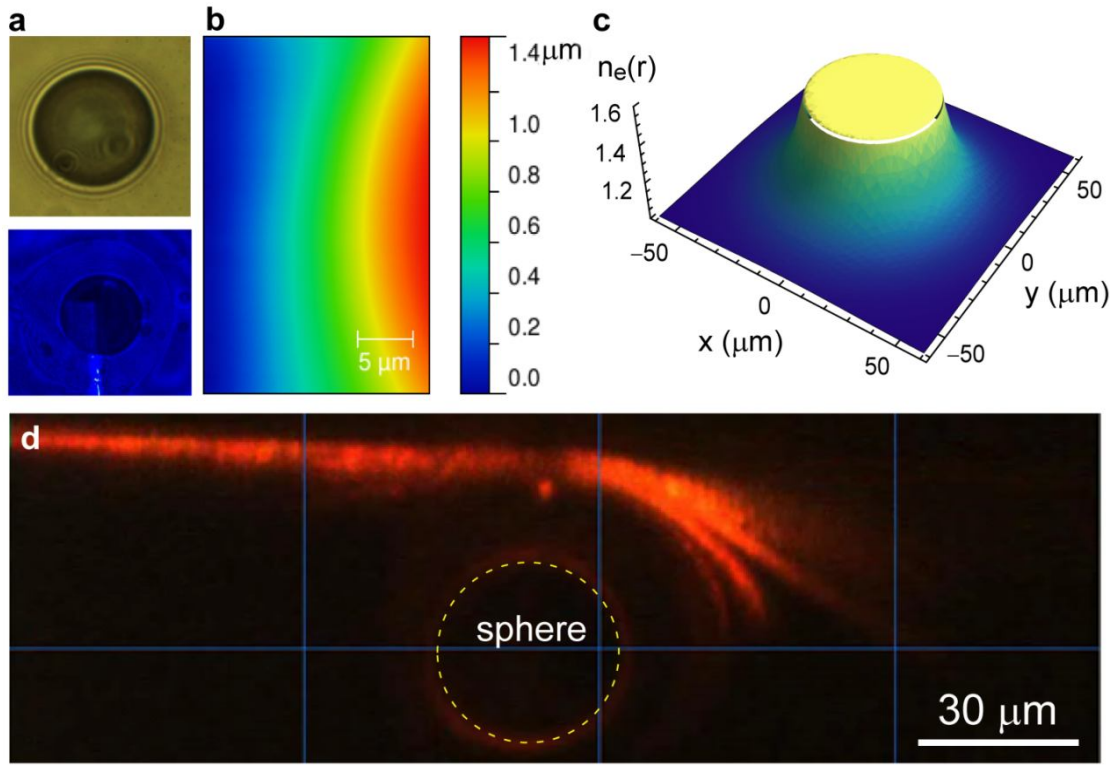

**Figure 2. Structural and optical measurements of the sample.** (a) The interference pattern around the microsphere illuminated by a white light (top) and blue light (bottom). (b) The surface profile of the PMMA layer measured with an atomic force microscopy (AFM). (c) The effective refractive index of the micro-structured waveguide is extracted showing strong power law dependence with the radial distance from the microsphere. (d) A particular example of light bending in close proximity to the microsphere. The incident beam is coupled into the waveguide using a diffraction grating drilled in the metal layer.

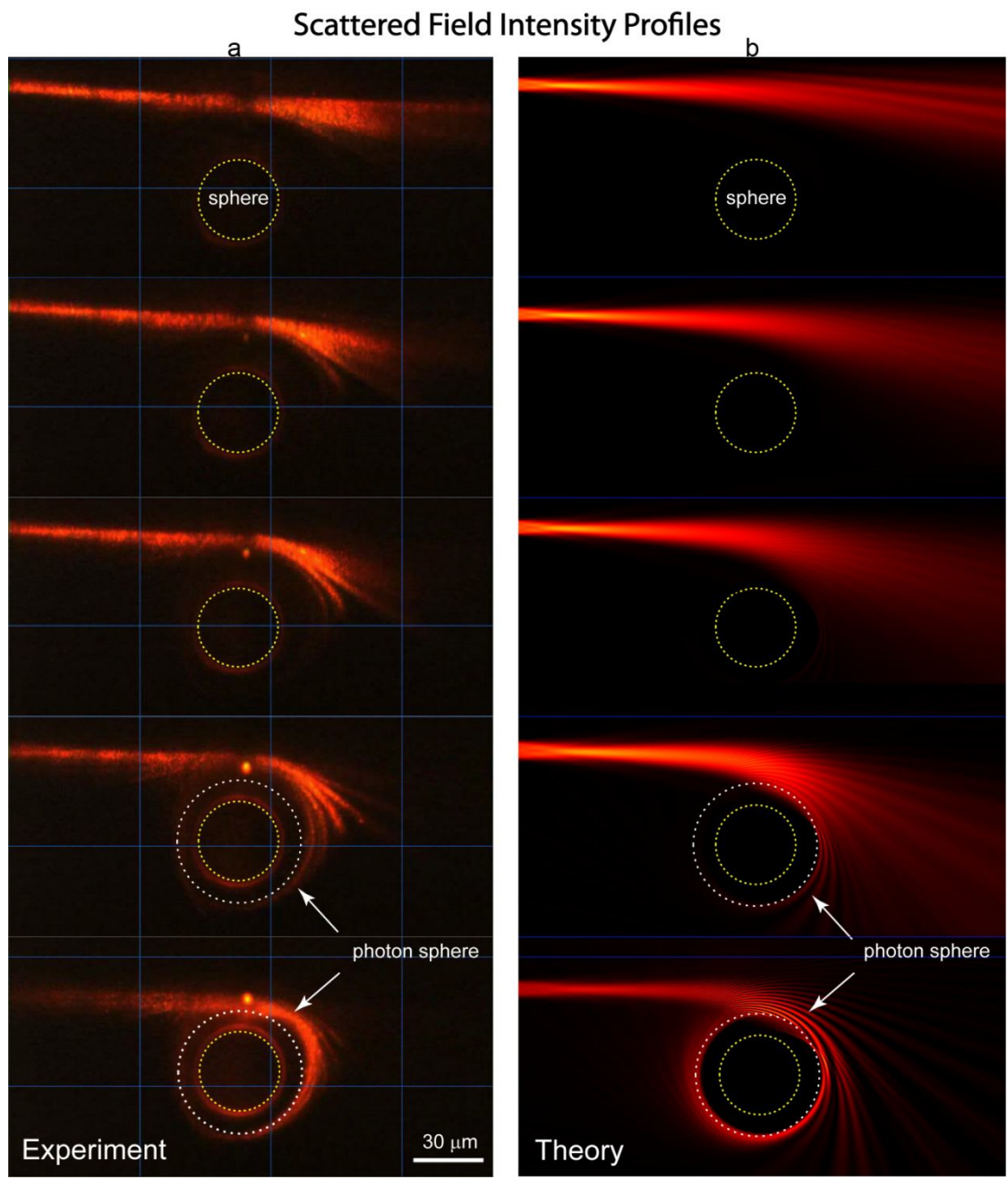

**Figure 3. The scattered field intensity around the micosphere** (a) observed in the experiment and (b) calculated using a full wave finite difference frequency domain (FDFD) electromagnetic code. In the calculations the effective refractive index is extracted from the experimental data. Starting from the top figures and going down the incident beam impact parameters is gradually decreased which results in strong increase in the beam deflection from its original path. At critical impact parameters (bottom two pictures) the light rays approach

the photon sphere with a fraction of the incident energy scattered array from the microsphere while the rest is captured around the microsphere.

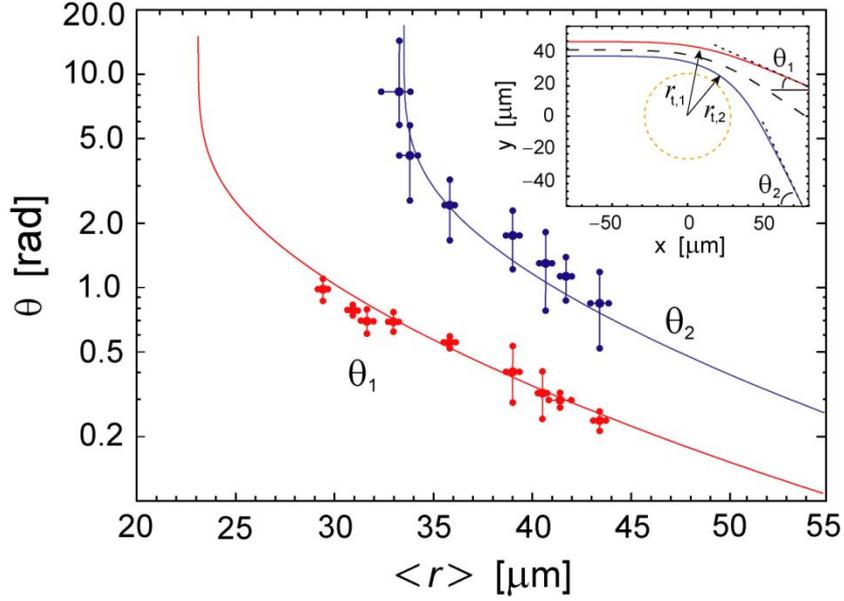

**Figure 4. The deflection angles measured in the experiment (dots) and calculated (red and blue lines) based on Eq. 3.** Due to the final width of the incident light beam two deflection angles $\theta_1$ and $\theta_2$, corresponding to the edges of the beam (at $1/e$ intensity), can be extracted unambiguously. The beam envelop is then represented by two points of closes approach $r_{t1}$ and $r_{t2}$ corresponding to the two deflection angles as shown in the figure insert and calculated using Eq. 2. For the purpose of presentation we depict the deflection angles as function of the geometrical average between the two turning distances: $\langle r \rangle = (r_{t1} + r_{t2})/2$. The error bars due to the experiment are also included. The experimental data closely marches the theory for all measured cases. A singularity in the deflection angle is observed for $r_{t2} \approx a = 28.5 \mu m$, corresponding to the photon sphere of our system.

*Supplementary information*

# Trapping light by mimicking gravitational lensing

C. Sheng[1], H. Liu[1,*], Y. Wang[1], S. N. Zhu[1], and D. A. Genov[2]

[1] National Laboratory of Solid State Microstructures & Department of Physics, Nanjing University, Nanjing 210093, People's Republic of China

[2] College of Engineering and Science, Louisiana Tech University, Ruston, Louisiana 71270, USA

*corresponding author: liuhui@nju.edu.cn

## I. Experimental measurements, fittings and effective permittivity

The poly-methyl-methacrylate (PMMA) thickness profiles were measured using twodifferent methods: (a) direct Atomic Force Microscopy (AFM), and (b) interference data.The results are shown in Figure 1. The measured film thickness at large distances is then fitted with a power law function of the type

$$h(r) \approx h_\infty \left(1 + \left(\frac{R}{r}\right)^s\right) \qquad (1)$$

where $h_\infty$ is the film thickness at infinity. The experimental data are presented in Figure 1 while the fitting parameters are listed in Table 1. Both the AFM data and interference data show almost the same far-distance behaviors with similar power law exponents of about $s = 4$. The effective refractive waveguide index (permittivity) is obtained by considering the structured waveguide geometry consisting of Air/PMMA/Silver/SiO$_2$ multilayer stack. For transverse magnetic (TM) waves, the dispersion relationship is given as:

$$\exp[2ik_2h] = \frac{1 + r_{32}r_{43}\exp[2ik_3t]}{r_{12}r_{32} + r_{12}r_{43}\exp[2ik_3t]} \quad (2)$$

where $r_{ik} = (\eta_k - \eta_i)/(\eta_k + \eta_i)$ are the reflection coefficients, $t$ is the metal layer thickness, the specific impedances and transversal wave vectors are $\eta_i = (n_i^2 - n_e^2)^{1/2}/n_i^2$ and $k_i = (\omega/c)(n_i^2 - n_e^2)^{1/2}$, respectively, where $n_e$ is the effective waveguide index and $n_i$ are the refractive indexes of each media ((1)-air, (2)-PMMA, (3)-Silver, and (4)-Glass). Note that Eq. (2) cannot be solved explicitly with respect to the effective index but it can be solved with respect to the film thickness giving

$$h = \frac{1}{k_2}\left(m\pi + \frac{1}{2i}\ln\left(\frac{1 + r_{32}r_{43}\exp[2ik_3t]}{r_{12}r_{32} + r_{12}r_{43}\exp[2ik_3t]}\right)\right) \quad (3)$$

where $m$ is the mode index. For sufficiently thick metal layers Eq. (3) can be significantly simplified as

$$h = \frac{1}{k_2}\left(m\pi + \frac{i}{2}\ln(r_{12}r_{32})\right) \quad (4)$$

The dispersion relationship for the first seven modes calculated using Eq. (3) and Eq. (4) are shown in Figure 1c. In the calculations we use: $n_{SiO_2} = 1.47$[1], $n_{PMMA} = 1.51$[2], and at the operation wavelength $\lambda = 405nm$, the silver index $n_{Ag} = 0.05 + i2.275$ is taken from Johnson&Christy[3]. The metal film thickness is set at $t = 50nm$. Clearly, the metal layer is sufficient thick to reduce the problem into a single dielectric waveguide on a top of a semi-infinite metal substrate. The estimated effective refractive index (permittivity) is then fitted to a power law function of the type

$$\varepsilon(r) = n^2(r) \approx n_\infty^2\left[1 + \left(\frac{a}{r}\right)^4\right] \quad (5)$$

where $n_\infty$ is the effective refractive index at infinity, and $a$ is the photon sphere radius (see below). The fitting parameters are included in Table 1.

| Data type | $h_\infty [\mu m]$ | $R [\mu m]$ | s | $a [\mu m]$ |
|---|---|---|---|---|
| AFM | 1.0515 | 26.542 | 4.042 | 28.5 |
| Interference | 1.071 | 26.43 | 4.245 | 27.96 |

**Table 1.** Experimental fitting parameters for the polymethylmethacrylate (PMMA) film thickness and effective refractive index of the $TM_{0,6}$ mode.

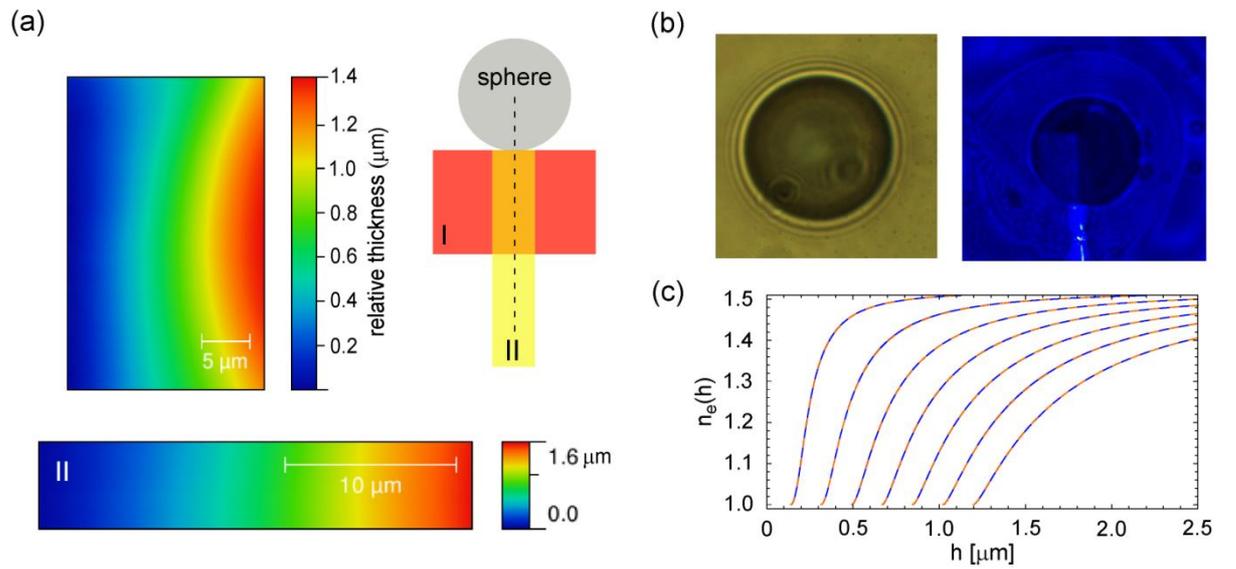

**Figure 1.** Film thickness measured using (a) atomic force microscope (AFM) data and (b) interference data. (c) The effective refraction index is calculated using Eq. (3) (blue solid line) and Eq. (4) (orange dashed line).

## II. Centrally-symmetric isotropic spacetimes

**Einstein field equations:** Here, we seek to identify acurved spacetime that canprovide optical effects similar to those observed in the experiment.We begin by considering a static centrally symmetric spacetime in homogeneous coordinates

$$ds^2 = e^{2\nu}c^2 dt^2 - e^{2\lambda}(dr^2 + r^2 d\Omega^2) \quad (6)$$

where the metric functions $v$ and $\lambda$ depend on the radial coordinate only. The metric elements are solution of the Einstein field equations

$$G_{\mu\nu} = -T_{\mu\nu} = -\rho u_\mu u_\nu - p(u_\mu u_\nu - g_{\mu\nu})$$

where $u_\mu$ are the 4-velocities and wehave adopted a system of units with $8\pi G = c = 1$.

Following a standard procedure[4] the governing equationsare as follows:

$$\rho = -\frac{e^{-2\lambda}}{r}(2r\lambda'' + \lambda'(4 + r\lambda'))$$
$$p = \frac{e^{-2\lambda}}{r}(2v'(1 + r\lambda') + \lambda'(2 + r\lambda')) \quad (7)$$
$$(\rho + p)v' + p' = 0$$

The above equations are solved either by providing the equation of state $p = p(\rho)$, or by using a generating function. Here we will rely on the later by introducing the effective refractive indexof the metric

$$n^2 = e^{2(\lambda-v)}, \quad \lambda(r) = v(r) + \ln[n(r)].$$

Working with the non-dimensional radial coordinate$\frac{r}{a} \to r$ we obtain an analytical solution of Eq.7 with the metric elements given as

$$e^{2\lambda} = -\frac{A}{1+r^4}\sec^2(\varphi + \delta), \quad e^{2v} = -\frac{Ar^4}{(1+r^4)^2}\sec^2(\varphi + \delta) \quad (8)$$

where$\varphi = \sqrt{5/2}\cot^{-1}(r^2)$, $\delta \in \mathbb{R}$ and$A \in \mathbb{R}$ are the integration constants.All unique solutions are obtained for $\delta \in \left(-\frac{\pi}{2}, \frac{\pi}{2}\right)$.

**Mass-density and pressure:** The mass-density and pressures are obtained from Eqs. 7 and 8

$$\rho = \frac{6\sin[\varphi + \delta](5r^2\sin[\varphi + \delta] + \sqrt{10}(r^4 - 1)\cos[\varphi + \delta])}{A(1 + r^4)}$$

$$p = -\frac{2 + 9r^4 + 2r^8 + (2 - 21r^4 + 2r^8)\cos[2(\varphi + \delta)] + 4\sqrt{10}r^2(r^4 - 1)\sin[2(\varphi + \delta)]}{Ar^2(1 + r^4)}$$

*Special case - finite pressure and mass density at large distances:*The only solution with finite positive pressure and density at large distances is obtained for$\delta = \pm\frac{\pi}{2}$, with

the metricgiven as

$$ds^2 = A\left(\frac{\csc(\varphi)}{r^2 n^2(r)}\right)^2 [-dt^2 + n^2(r)d\vec{x}^2] \qquad (9)$$

where $A > 0$. The metrics is finite and asymptotically flat at large distances(see Fig. 2).

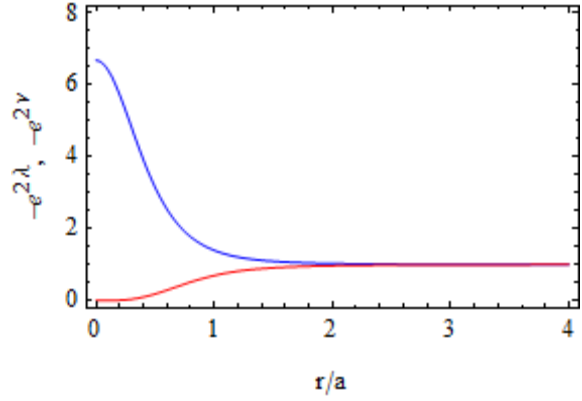

**Figure 2**The metric elements for $\delta = \pm\frac{\pi}{2}$ and $A = 5/2$.

The pressure and mass-density are

$$\rho = 3\frac{5r^2 + 5r^2\cos[2\varphi] + \sqrt{10}(1-r^4)\sin[2\varphi]}{A(1+r^4)}$$

$$p = \frac{(2 - 21r^4 + 2r^8)\cos[2\varphi] - 4\sqrt{10}r^2(1-r^4)\sin[2\varphi] - 2 - 9r^4 - 2r^8}{Ar^2(1+r^4)}$$

Both, the pressure and the mass density are positive for $r \geq r_c = a\sqrt{\cot(\pi/\sqrt{10})} = 0.8071a$ which is the range where the effective refractive index fit(Eq. 5) is valid. Clearly, any physically meaningful solution that covers the entire space must be formed by switching the metrics Eq. 9 with some other interior solution that is valid for $r < a$. At large distances we have

$$\rho = \frac{15}{A}r^{-6} + O(r^{-10}), \quad p = \frac{3}{2A}r^{-10} + O(r^{-14})$$

which corresponds to an equation of state $p = k\rho^{5/3} = k\rho^{1+1/n}$, or apolytrop with n = 3/2. Such equation of state is typical for degenerate star cores including degenerated neutron stars, red giants, white and brown dwarfs.

**Geodesic trajectories:** Here, we consider the trajectories of massive/massless

particles in the spacetime described by the metrics Eq. 9 which we now recast as

$$ds^2 = g_{00}(r)\bigl[-dt^2 + n^2(r)(dr^2 + r^2 d\Omega^2)\bigr] \quad (10)$$

The Lagrangian $\mathcal{L}$ and the Euler-Lagrange equations for a ray motion restricted to the $\theta = \pi/2$ plane follows

$$\mathcal{L} = \kappa = \frac{1}{2}\left(\frac{ds}{d\tau}\right)^2 = \frac{1}{2}g_{00}(r)[-\dot{t}^2 + n^2(r)(\dot{r}^2 + r^2\dot{\varphi}^2)]$$
$$\dot{\varphi} = \frac{d\varphi}{d\tau} = \frac{a}{g_{00}(r)n^2(r)r^2}, \quad \dot{t} = \frac{dt}{d\tau} = \frac{a}{b g_{00}(r)} \quad (11)$$

where $a$ and $b$ are constants of motion related to the total energy and angular momentum of the particle, $\tau$ is an affine parameter and $\kappa = 1/0$ for massive/massless particle. Using the azimuthal angle $\varphi$ as the parameter of the trajectory, the equation of motion (first integral) for a massless particle is obtained from Eq. 11

$$\left(\frac{dr}{d\varphi}\right)^2 = \frac{n^2(r)r^4}{b^2} - r^2 \quad (12)$$

**Deflection angle:** Here, we seek to obtain an analytical solution of Eq. 12 with the use of the experimentally measured effective refractive index Eq. 5. It is convenient to use the inverse radial coordinate $u = a/r$, and Eq. (12) now reads

$$\left(\frac{du}{d\varphi}\right)^2 = q^2(1 + u^4) - u^2 \quad (13)$$

where $q = a/b$ is a constant. This equation has both implicit and explicit solutions in the form

$$\varphi(u) = \varphi_0 + (1 + u_t^4)^{1/2} F\left(\sin^{-1}\left(\frac{u}{u_t}\right)\Big| u_t^4\right)$$
$$u(\varphi) = u_0 + u_t \operatorname{sn}\left(\frac{q(\varphi - \varphi_0)}{u_t}\Big| u_t^4\right) \quad (14)$$

where $\varphi_0$ is the angle of incidence, $u_0$ is the initial position, $F$ is the elliptic integral of the first kind, and sn is the Jacobi elliptic function. The solution depends on the external turning point,

$$u_t = \frac{a}{r_t} = \frac{1}{q\sqrt{2}}\left(1 - \sqrt{1 - 4q^4}\right)^{\frac{1}{2}} = (b/b_c)\left(1 - \sqrt{1 - (b_c/b)^4}\right)^{1/2} \quad (15)$$

From Eq. (15) it follows that a turning point $u_t$ (position of closest approach) exist for in-falling rays only if $q \leq 1/\sqrt{2}$ or the impact parameter is larger than the critical value $b \geq b_c = a\sqrt{2}$. The position of the photon sphere is at $r_{ph} = a$ and the internal turning point is $1/u_t$. The turning points and the complete phase space diagram are shown in Fig. 3. Finally, the deflection angle is obtained from Eq. (14) and Eq. (15) and depends on the turning point or equivalently the impact parameter

$$\theta = 2K[u_t^4]\sqrt{1 + u_t^4} - \pi \quad (16)$$

where $K$ is the complete elliptical integral of the first kind. If the incident ray is far from the photon sphere ($b \gg b_c$), then $q \ll 1$ and we can simplify Eq. (16) to obtain

$$\theta \to \frac{3\pi q^4}{4} = 3\pi\left(\frac{b_c}{2b}\right)^4 \quad (17)$$

For rays with impact parameters close to the critical ($b \approx b_c$) the deflection angle experiences a logarithmic singularity, which is easily obtained by expanding Eq. (16)

$$\theta \to \frac{1}{\sqrt{2}}\ln\left(\frac{16}{1 - b_c/b}\right) - \pi \quad (18)$$

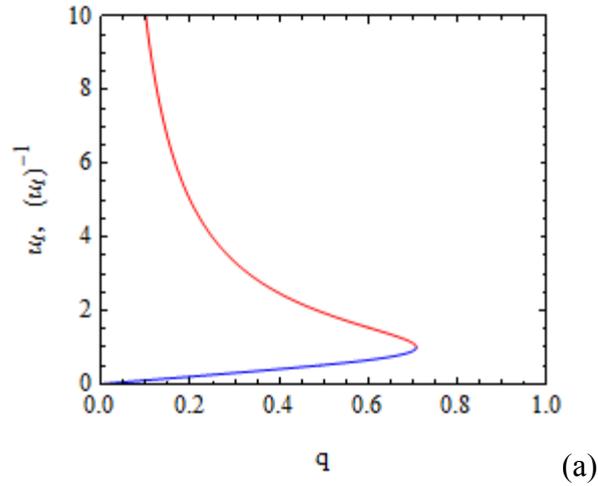

(a)

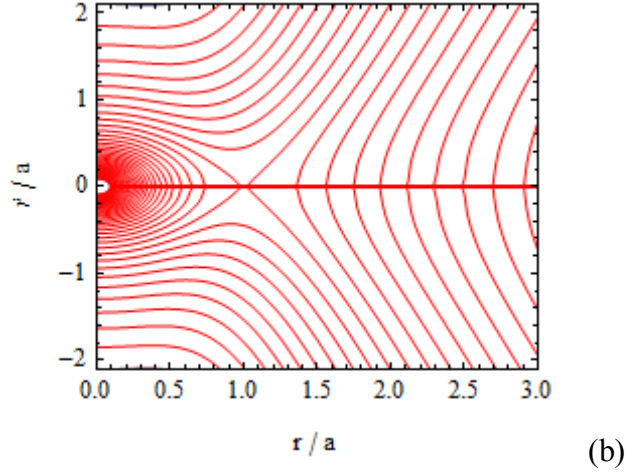

(b)

**Figure 3**(a)The external (blue) and internal (red)turning points, (b) phase space diagram corresponding to the experimentally fitted effective refractive index Eq. 5.

**Scattering and capture cross-lengths:** Eqs. (15) and (16), can be used to obtain the scattering differential cross-length

$$\frac{d\sigma(\theta)}{d\theta} = \left|\frac{db}{d\theta}\right| = \left|\frac{db}{du_t}\bigg/\frac{d\theta}{du_t}\right| = \frac{a}{4u_t}\left|\frac{(1-u_t^4)^2}{(1+u_t^4)E[u_t^4] - (1-u_t^4)K[u_t^4]}\right| \quad (19)$$

For small angles we have

$$\frac{d\sigma(\theta)}{d\theta} \approx \frac{a(3\pi)^{1/4}}{4\sqrt{2}\theta^{5/4}} \quad (20)$$

The total capture cross-length is $\sigma_t = 2b_c = 2a\sqrt{2}$, whereas the total scattering cross-length is infinite due to the long range interactions.

**Lens equation:** The long-distance result Eq. (17) allows writing an equivalent lens equation in the form

$$\frac{1}{s_1} + \frac{1}{s_2} = \frac{1}{f} \quad (21)$$

where $f = 4a/3\pi u_t^5$ is the equivalent "focal" length which depends on the distance of closest approach $r_t = a/u_t$ or the Einstein ring radius; and $s_1$ and $s_2$ are the distances to the center of the source and image, respectively.